\documentclass[twocolumn,english]{revtex4}
\usepackage[T1]{fontenc}
\usepackage[latin9]{inputenc}
\usepackage{float}
\usepackage{textcomp}
\usepackage{amsmath}
\usepackage{graphicx}



\usepackage{babel}

\begin{document}

\title{Symmetric lens with extended depth of focus}

\author{Sung Nae Cho  }

\email{sungnae.cho@samsung.com}

\affiliation{MEMS \& Packaging Group, Samsung Advanced Institute of Technology,
Mt. 14-1 Nongseo-dong Giheung-gu, Yongin-si Gyeonggi-do, 446-712,
South Korea }

\date{Prepared 17 July 2008 }

\begin{abstract}
The lens surface profile is derived based on the instantaneous focal
length versus the lens radius data. The lens design based on instantaneous
focal length versus the lens radius data has many useful applications
in software assisted image focusing technology.
\end{abstract}


\maketitle

\section{Introduction}

The software assisted image focusing is an emerging technology which
is expected to replace the traditional methods of autofocusing in
image manipulation devices, such as digital cameras and mobile phones.
Unlike the traditional methods, where the focusing of images is done
by mechanically movable parts, the software assisted technology produces
focused images by processing it through specialized image reconstruction
algorithm. The transition from mechanical to software assisted image
focusing can be attributed to the (1) demand for thinner and lighter
products by customers, and (2) the advancements in manufacturing process
for faster and more power efficient digital signal processors. With
autofocusing by mechanically movable parts, the demand for thinner
and lighter products is becoming a top hurdle for manufacturing process.
On the other hand, the advancements in more power efficient and faster
digital signal processors make software assisted image focusing technology
ideal for satisfying customer's demand for thinner and lighter image
developing products such as digital cameras and mobile phones to name
a few. 

At the heart of software assisted image autofocusing technology is
the specialized image reconstruction algorithm permanently coded into
the built in digital signal processor. The actual layout of the code
base for image reconstruction algorithm varies among different manufacturers
and many manufacturers do not disclose their algorithms to public
as they constitute a trade secret. The image reconstruction algorithm
can be codified based on instantaneous focal length versus the lens
radius data\citet{Getman}. Once this specialized image reconstruction
algorithm is adopted for the system, a lens must be designed so that
its output matches the instantaneous focal length versus the lens
radius data, which information was assumed and used as input to the
image reconstruction code base.

In this work, a formula for the lens surface profile is presented.
The derivation of lens surface profile is solely based on the instantaneous
focal length versus the lens radius data; and therefore, the result
is expected to find useful applications in software assisted image
focusing technologies.

\section{Instantaneous focal length data}

\begin{figure}[H]
\begin{centering}
\includegraphics[scale=0.65]{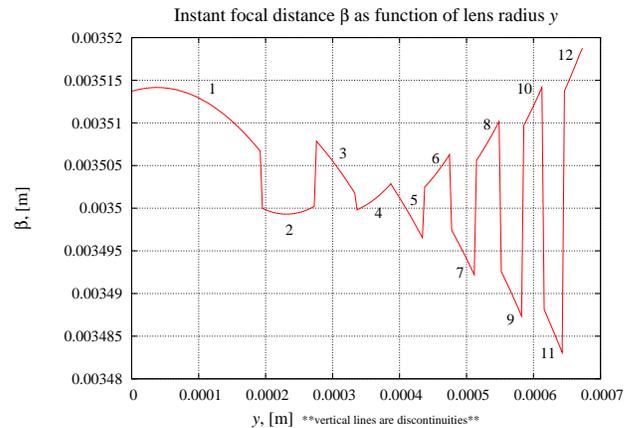}
\par\end{centering}

\caption{\label{fig:lens-data-0} Instantaneous focal length versus the lens
radius, where both are measured in meters. The normal incidence is
assumed for the incoming light rays. }

\end{figure}

Alexander and Lukyanov have recently filed for a patent which deals
with image reconstruction algorithm with applications in software
assisted image focusing technology. In their proposal, they claim
to have obtained an optimal image processing solution, which is expected
to be a significant improvement over the predecessor\citet{Portney}.
Behind their optimization is the instantaneous focal length versus
the lens radius data illustrated in Fig. \ref{fig:lens-data-0}, which
result assumes a normal incidence for the incident light waves. In
the figure, $\beta$ denotes the instantaneous focal length and $y$
is the lens radius. Each of the twelve segmented curves can be represented
by the quadratic polynomial \begin{align}
\beta_{i}\left(y\right) & =a_{i}y^{2}+b_{i}y+c_{i},\label{eq:beta-i}\end{align}
 with coefficients $\left(a_{i},b_{i},c_{i}\right)$given by \[
a_{1}=-313.07,\; b_{1}=0.0235,\; c_{1}=0.0035137034,\]
 \[
a_{2}=534.53,\; b_{2}=-0.2472,\; c_{2}=0.003527877626,\]
 \[
a_{3}=-309.02,\; b_{3}=0.0818,\; c_{3}=0.0035088062,\]
 \[
a_{4}=536.05,\; b_{4}=-0.3275,\; c_{4}=0.0035493232,\]
 \[
a_{5}=-306.12,\; b_{5}=0.1182,\; c_{5}=0.003502912672,\]
 \[
a_{6}=539.03,\; b_{6}=-0.3891,\; c_{6}=0.003569538239,\]
 \[
a_{7}=-303.68,\; b_{7}=0.1463,\; c_{7}=0.003496845208,\]
 \[
a_{8}=542.21,\; b_{8}=-0.4417,\; c_{8}=0.003589312176,\]
 \[
a_{9}=-301.27,\; b_{9}=0.1695,\; c_{9}=0.00349080193,\]
 \[
a_{10}=545.96,\; b_{10}=-0.4895,\; c_{10}=0.003609151039,\]
 \[
a_{11}=-298.81,\; b_{11}=0.1893,\; c_{11}=0.003484870978,\]
 \[
a_{12}=179.08,\; b_{12}=-0.0474,\; c_{12}=0.003469596542,\]
 where the subscript $i$ denotes the $i\textup{th}$ curved segment
in Fig. \ref{fig:lens-data-0}. The curve fitting was done by linear
regression. The physical lens, whose output satisfies the instantaneous
focal length versus the lens radius data defined in Fig. \ref{fig:lens-data-0},
is one of the variants of lens with extended depth of focus\citet{Bradburn,Bradburn2,Forster}.
With Eq. (\ref{eq:beta-i}), I shall solve for the lens surface profile
whose output matches the instantaneous focal length versus the lens
radius data defined in Fig. \ref{fig:lens-data-0}.

\section{The lens surface equation}

\subsection{Derivation}

\begin{figure}[H]
\begin{centering}
\includegraphics[scale=0.45]{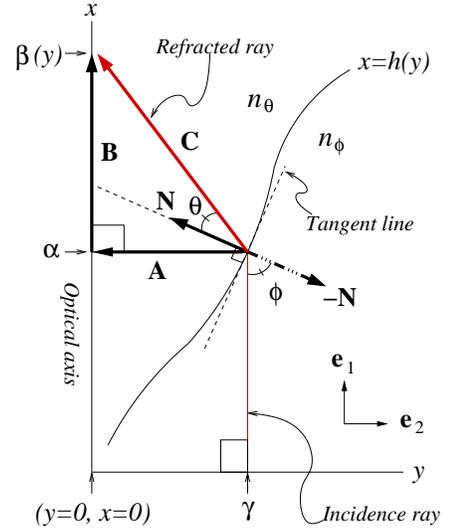}
\par\end{centering}

\caption{\label{fig:Law_of_refraction} Illustration of Snell's law.}

\end{figure}

When a ray of light passes across media of different refractive indices,
its path is governed by the Snell's law, \begin{equation}
n_{\phi}\sin\phi=n_{\theta}\sin\theta,\label{eq:Snell-law}\end{equation}
 as illustrated in Fig.  \ref{fig:Law_of_refraction}. Here, $n_{\phi}\equiv n_{\phi}\left(\omega\right)$
and $n_{\theta}\equiv n_{\theta}\left(\omega\right)$ are frequency
dependent refractive indices with $\omega$ denoting the angular frequency
of the light. The parameters $\phi$ and $\theta$ represent the angle
of incidence and angle of refraction, respectively.  

If $\mathbf{N}$ denotes the normal vector to the local point $y=\gamma$
on the curve $x=h\left(y\right),$ then it can be shown \[
\left\Vert -\mathbf{N}\times\left(-\mathbf{e}_{1}\right)\right\Vert =\left\Vert -\mathbf{N}\right\Vert \left\Vert -\mathbf{e}_{1}\right\Vert \sin\phi=N\sin\phi\]
 and the expression for $\sin\phi$ becomes \begin{eqnarray}
\sin\phi=\frac{\left\Vert \mathbf{N}\times\mathbf{e}_{1}\right\Vert }{N}, &  & N\equiv\left\Vert \mathbf{N}\right\Vert ,\label{eq:sin-of-phi}\end{eqnarray}
 where $\mathbf{e}_{1}$ is the unit basis for the $x$ axis. 

Similarly, the expression for $\sin\theta$ may be obtained by considering
vectors $\mathbf{A},$ $\mathbf{B},$ and $\mathbf{C}$ of Fig.  \ref{fig:Law_of_refraction}.
The vectors $\mathbf{A},$ $\mathbf{B},$ and $\mathbf{C}$ satisfy
the relation, \begin{equation}
\mathbf{A}+\mathbf{B}=\mathbf{C}.\label{eq:C_pre1}\end{equation}
 In explicit form, vectors $\mathbf{A}$ and $\mathbf{B}$ are defined
as \begin{equation}
\mathbf{A}=-\gamma\mathbf{e}_{2},\quad\mathbf{B}=\left(\beta-\alpha\right)\mathbf{e}_{1},\label{eq:C_pre2}\end{equation}
 where $\mathbf{e}_{2}$ is the unit basis for the $y$ axis. With
Eqs. (\ref{eq:C_pre1}) and (\ref{eq:C_pre2}), the vector $\mathbf{C}$
becomes \begin{equation}
\mathbf{C}=\left(\beta-\alpha\right)\mathbf{e}_{1}-\gamma\mathbf{e}_{2}.\label{eq:C}\end{equation}
 The vector cross product $\mathbf{N}\times\mathbf{C}$ is given by
\[
\mathbf{N}\times\mathbf{C}=\left(\beta-\alpha\right)\mathbf{N}\times\mathbf{e}_{1}-\gamma\mathbf{N}\times\mathbf{e}_{2}\]
 and its magnitude becomes \begin{equation}
\left\Vert \mathbf{N}\times\mathbf{C}\right\Vert =\left\Vert \left(\beta-\alpha\right)\mathbf{N}\times\mathbf{e}_{1}-\gamma\mathbf{N}\times\mathbf{e}_{2}\right\Vert =NC\sin\theta,\label{eq:sin-of-theta-pre}\end{equation}
 where $N\equiv\left\Vert \mathbf{N}\right\Vert $ and $C=\left\Vert \mathbf{C}\right\Vert .$
Utilizing Eq. (\ref{eq:C}), $C$ may be expressed as  \[
C=\left(\mathbf{C}\cdot\mathbf{C}\right)^{1/2}=\left[\left(\beta-\alpha\right)^{2}+\gamma^{2}\right]^{1/2}\]
 and the Eq. (\ref{eq:sin-of-theta-pre}) is solved for $\sin\theta$
to yield \begin{equation}
\sin\theta=\frac{\left\Vert \left(\beta-\alpha\right)\mathbf{N}\times\mathbf{e}_{1}-\gamma\mathbf{N}\times\mathbf{e}_{2}\right\Vert }{N\left[\left(\beta-\alpha\right)^{2}+\gamma^{2}\right]^{1/2}}.\label{eq:sin-of-theta}\end{equation}

Insertion of Eqs. (\ref{eq:sin-of-phi}) and (\ref{eq:sin-of-theta})
into the Snell's law of Eq. (\ref{eq:Snell-law}) gives \begin{equation}
\frac{n_{\phi}}{n_{\theta}}=\frac{\left\Vert \left(\beta-\alpha\right)\mathbf{N}\times\mathbf{e}_{1}-\gamma\mathbf{N}\times\mathbf{e}_{2}\right\Vert }{\left\Vert \mathbf{N}\times\mathbf{e}_{1}\right\Vert \left[\left(\beta-\alpha\right)^{2}+\gamma^{2}\right]^{1/2}}.\label{eq:Snell-alt-1}\end{equation}
 By definition, the normal vector $\mathbf{N}$ satisfies the relation,
 \[
g\left(x,y\right)=x-h\left(y\right),\]
 where $g\left(x,y\right)$ is a function whose gradient gives $\mathbf{N},$
\[
\mathbf{N}=\nabla g=\frac{\partial g}{\partial x}\mathbf{e}_{1}+\frac{\partial g}{\partial y}\mathbf{e}_{2}=\mathbf{e}_{1}-\frac{\partial h}{\partial y}\mathbf{e}_{2}.\]
 Because $\mathbf{N}$ is the normal vector at the location $\left(x=\alpha,y=\gamma\right),$
I write \begin{equation}
\mathbf{N}=\mathbf{e}_{1}-\left.\frac{\partial h}{\partial y}\right|_{y=\gamma}\mathbf{e}_{2}.\label{eq:N-explicit-xa}\end{equation}
 The following vector cross products are valid, \begin{align*}
\mathbf{N}\times\mathbf{e}_{1} & =\mathbf{e}_{1}\times\mathbf{e}_{1}-\left.\frac{\partial h}{\partial y}\right|_{y=\gamma}\mathbf{e}_{2}\times\mathbf{e}_{1},\\
\mathbf{N}\times\mathbf{e}_{2} & =\mathbf{e}_{1}\times\mathbf{e}_{2}-\left.\frac{\partial h}{\partial y}\right|_{y=\gamma}\mathbf{e}_{2}\times\mathbf{e}_{2},\end{align*}
 where Eq. (\ref{eq:N-explicit-xa}) was used to replace $\mathbf{N}.$
Since $\mathbf{e}_{1}\times\mathbf{e}_{1}=\mathbf{e}_{2}\times\mathbf{e}_{2}=0,$
the previous relations reduce to \begin{equation}
\mathbf{N}\times\mathbf{e}_{1}=\left.\frac{\partial h}{\partial y}\right|_{y=\gamma}\mathbf{e}_{3},\quad\mathbf{N}\times\mathbf{e}_{2}=\mathbf{e}_{3},\label{eq:N-cross-ei}\end{equation}
 where $\mathbf{e}_{3}$ is the unit basis for the $z$ axis of which
satisfies the relation,  \[
\mathbf{e}_{1}\times\mathbf{e}_{2}=\mathbf{e}_{3},\quad\mathbf{e}_{2}\times\mathbf{e}_{1}=-\mathbf{e}_{3}.\]
 Insertion of Eq. (\ref{eq:N-cross-ei}) into Eq. (\ref{eq:Snell-alt-1})
gives \[
\frac{n_{\phi}}{n_{\theta}}=\frac{\left(\beta-\alpha\right)\left.\frac{\partial h}{\partial y}\right|_{y=\gamma}-\gamma}{\left.\frac{\partial h}{\partial y}\right|_{y=\gamma}\left[\left(\beta-\alpha\right)^{2}+\gamma^{2}\right]^{1/2}},\]
 which expression can be rearranged to yield \begin{equation}
\left.\frac{\partial h}{\partial y}\right|_{y=\gamma}=\frac{\gamma}{\beta-\alpha-\frac{n_{\phi}}{n_{\theta}}\left[\left(\beta-\alpha\right)^{2}+\gamma^{2}\right]^{1/2}},\label{eq:Snell-alt-2}\end{equation}
 where $\alpha$ and $\gamma$ are constants of which are depicted
in Fig. \ref{fig:Law_of_refraction}. 

For Alexander and Lukyanov's optical element, the instantaneous focal
function $\beta\equiv\beta\left(y\right)$ in Eq. (\ref{eq:Snell-alt-2})
is as defined in Fig. \ref{fig:lens-data-0}. The $\gamma$ for the
$y$ axis is not anything special, of course. Any $y$ belonging to
the domain of $h$ satisfies the Eq. (\ref{eq:Snell-alt-2}). The
generalization of Eq. (\ref{eq:Snell-alt-2}) for all $y$ belonging
to the domain of $h$ is done by making the following replacements:
\[
\alpha\rightarrow x,\quad\gamma\rightarrow y,\quad\left.\frac{\partial h}{\partial y}\right|_{y=\gamma}\rightarrow\frac{\partial h}{\partial y}=\frac{dx}{dy}.\]
 With these replacements, Eq. (\ref{eq:Snell-alt-2}) gets re-expressed
in form as \begin{equation}
\frac{dx}{dy}=\frac{y}{\beta-x-\frac{n_{\phi}}{n_{\theta}}\left[\left(\beta-x\right)^{2}+y^{2}\right]^{1/2}}.\label{eq:dxdy0}\end{equation}

How is the instantaneous focal function, $\beta,$ restricted? The
$\beta$ in Eq. (\ref{eq:dxdy0}) is restricted so that the expression
for $dx/dy$ does not blow up. Equation (\ref{eq:dxdy0}) is well
behaved if and only if the denominator satisfies the condition, \[
\beta-x-\frac{n_{\phi}}{n_{\theta}}\left[\left(\beta-x\right)^{2}+y^{2}\right]^{1/2}\neq0.\]
 To solve for $\beta,$ I shall first rearrange the previous expression
to get \[
\beta-x\neq\frac{n_{\phi}}{n_{\theta}}\left[\left(\beta-x\right)^{2}+y^{2}\right]^{1/2}.\]
Squaring both sides, \[
\left(\beta-x\right)^{2}\neq\frac{n_{\phi}^{2}}{n_{\theta}^{2}}\left(\beta-x\right)^{2}+\frac{n_{\phi}^{2}}{n_{\theta}^{2}}y^{2},\]
 and regrouping the terms, I find \[
\left(\beta-x\right)^{2}\left(1-\frac{n_{\phi}^{2}}{n_{\theta}^{2}}\right)\neq\frac{n_{\phi}^{2}}{n_{\theta}^{2}}y^{2}.\]
 The resulting expression can be solved for $\beta$ to yield \begin{equation}
\beta\neq x\pm\frac{n_{\phi}y}{\sqrt{n_{\theta}^{2}-n_{\phi}^{2}}}.\label{eq:beta-restriction}\end{equation}
 Equation (\ref{eq:beta-restriction}) defines the restriction for
the instantaneous focal function.

\subsection{Lens surface profile}

The profile of axially symmetric lens about its optical axis is obtained
by solving the initial-value differential equation, Eq. (\ref{eq:dxdy0}),
\[
\frac{dx}{dy}=\frac{y}{\beta-x-\frac{n_{\phi}}{n_{\theta}}\left[\left(\beta-x\right)^{2}+y^{2}\right]^{1/2}},\quad x\left(y_{\textup{0}}\right)=x_{\textup{0}},\]
 where $x\left(y_{\textup{0}}\right)=x_{\textup{0}}$ is the initial
condition to be specified and the instantaneous focal function $\beta$
satisfies the constrain defined in Eq. (\ref{eq:beta-restriction}).
Without loss of generality, one may choose $x\left(y=y_{\textup{0}}=0\right)=0$
for the initial condition and the lens profile satisfies the differential
equation, \begin{align}
\frac{dx}{dy} & =\frac{y}{\beta_{i}-x-\frac{n_{\phi}}{n_{\theta}}\left[\left(\beta_{i}-x\right)^{2}+y^{2}\right]^{1/2}},\nonumber \\
\label{eq:ODE}\\ & x\left(0\right)=0,\quad\beta_{i}\neq x\pm\frac{n_{\phi}y}{\sqrt{n_{\theta}^{2}-n_{\phi}^{2}}},\nonumber \end{align}
 where the index $i$ in $\beta_{i}$ comes from the fact that the
input specification defined in Fig. \ref{fig:lens-data-0} is piece
wise continuous over range of $x.$ The domain for each $\beta_{i}$
is given by \[
\beta_{1}:\quad0\leq y\leq0.00019182692,\]
 \[
\beta_{2}:\quad0.00019519231\leq y\leq0.00027259615,\]
 \[
\beta_{3}:\;0.00027596154\leq y\leq0.00033317308,\]
 \[
\beta_{4}:\quad0.00033653846\leq y\leq0.00038701923,\]
 \[
\beta_{5}:\quad0.00039038462\leq y\leq0.00043413462,\]
 \[
\beta_{6}:\quad0.0004375\leq y\leq0.00047451923,\]
 \[
\beta_{7}:\quad0.00047788462\leq y\leq0.00051153846,\]
 \[
\beta_{8}:\quad0.00051490385\leq y\leq0.00054855769,\]
 \[
\beta_{9}:\quad0.00055192308\leq y\leq0.00058221154,\]
 \[
\beta_{10}:\quad0.00058557692\leq y\leq0.0006125,\]
 \[
\beta_{11}:\quad0.00061586538\leq y\leq0.00064278846,\]
 \[
\beta_{12}:\quad0.00064615385\leq y\leq0.00067307692.\]

\begin{figure}[H]
\begin{centering}
\includegraphics[scale=0.65]{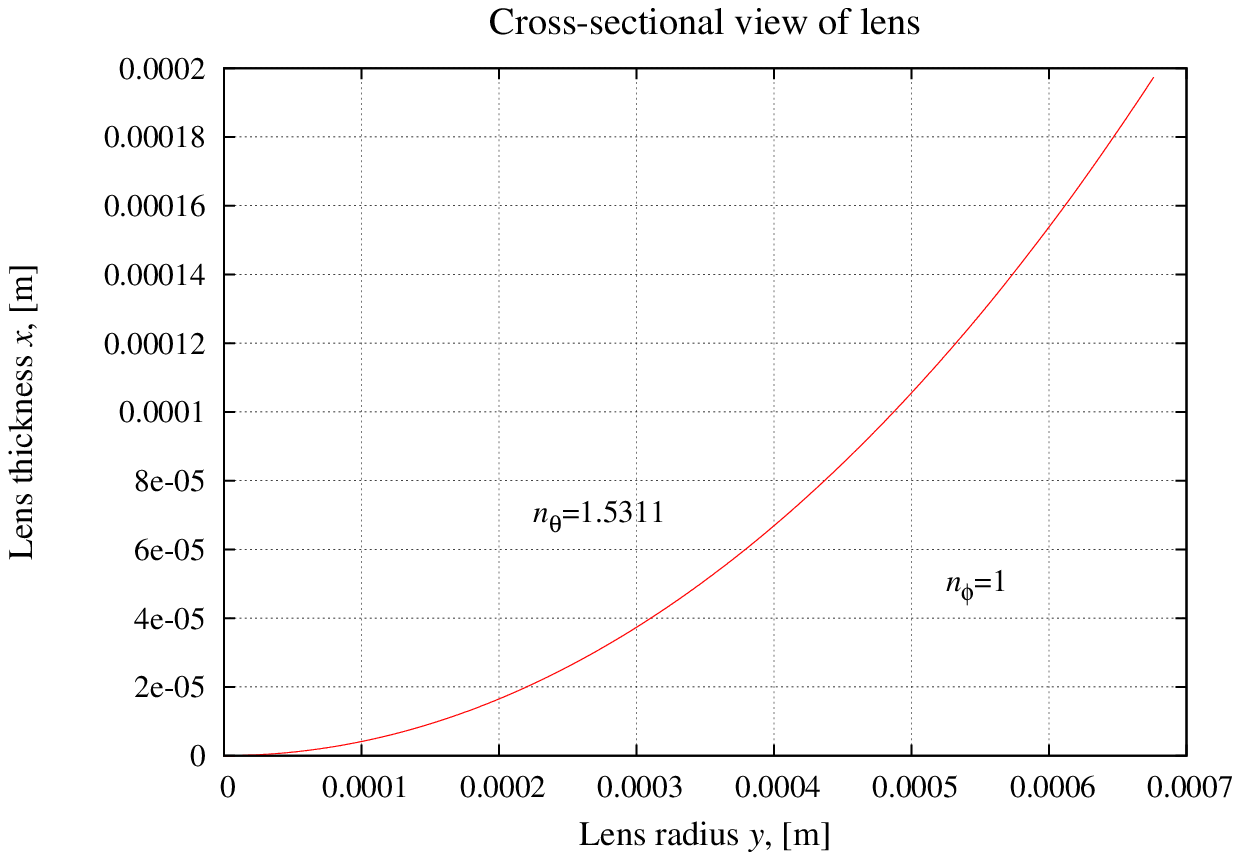}
\par\end{centering}

\caption{\label{fig:n-phi-air} The lens cross-section, $n_{\theta}>n_{\phi}.$}

\end{figure}

\begin{figure}[H]
\begin{centering}
\includegraphics[scale=0.45]{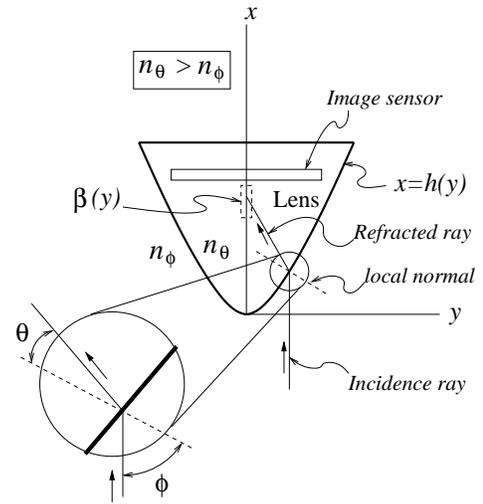}
\par\end{centering}

\caption{\label{fig:lens-with-sensor-np-air} Location of the  image sensor.}

\end{figure}

The differential equation (\ref{eq:ODE}) has been solved using the
Runge-Kutta method\citet{Derrick-Grossman}. The Runge-Kutta routine
has been coded in \textbf{FORTRAN 90} and the result for the case
where $n_{\phi}=1$ and $n_{\theta}=1.5311$ is provided in Fig. \ref{fig:n-phi-air}.
The physical lens may be obtained by revolving the curve about the
$x$ axis. Since the light ray is directed in the positive $x$ direction,
as illustrated in Fig. \ref{fig:lens-with-sensor-np-air}, it implies
that the image sensor should be embedded inside the lens for the case
where $n_{\theta}>n_{\phi}.$ 

\begin{figure}[H]
\begin{centering}
\includegraphics[scale=0.65]{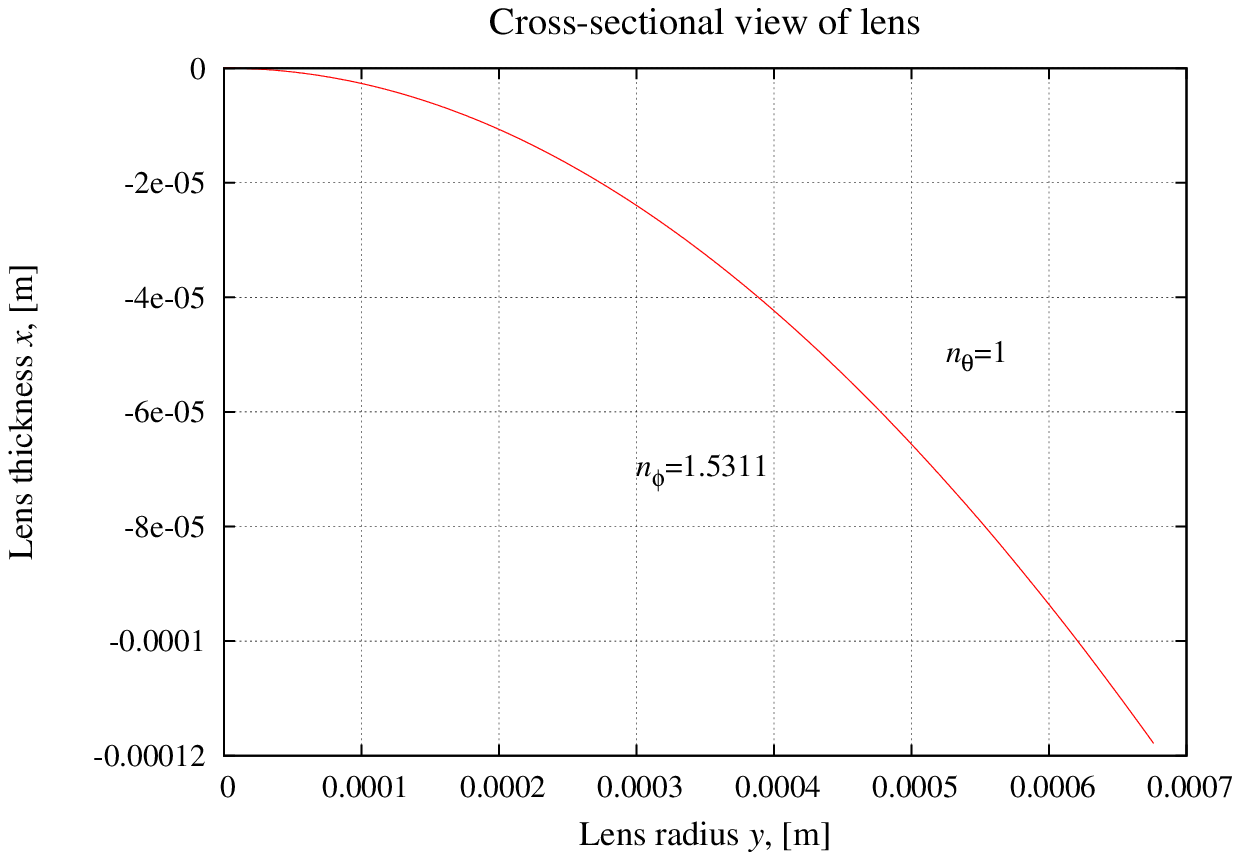}
\par\end{centering}

\caption{\label{fig:n-theta-air} The lens cross-section, $n_{\theta}<n_{\phi}.$}

\end{figure}

\begin{figure}[H]
\begin{centering}
\includegraphics[scale=0.45]{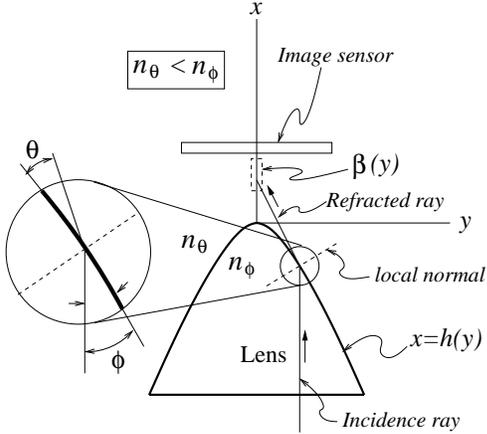}
\par\end{centering}

\caption{\label{fig:lens-with-sensor-nt-air} Location of the image sensor.}

\end{figure}

Reversing the values for two refractive indices, i.e., ($n_{\phi}=1.5311$
and $n_{\theta}=1$), the lens surface profile becomes as illustrated
in Fig. \ref{fig:n-theta-air}. Again, the physical lens may be obtained
by revolving the curve about the $x$ axis. Since the light ray is
directed in the positive $x$ direction, the case $n_{\theta}<n_{\phi}$
represents the situation where light is exiting the lens medium. For
this configuration, where $n_{\theta}<n_{\phi},$ the image sensor
should be placed external to the lens medium, as illustrated in Fig.
\ref{fig:lens-with-sensor-nt-air}.

The plots of lens surface cross-section, Figs. \ref{fig:n-phi-air}
and \ref{fig:n-theta-air}, deceptively portray as if lens surface
profile is represented by parabolic class of  curves. To show graphically
that this is not the case, the numerical data solutions obtained via
Runge-Kutta method to graph Figs. \ref{fig:n-phi-air} and \ref{fig:n-theta-air}
were linearly regressed to obtain \begin{align}
x_{1} & =a_{1}y^{6}+b_{1}y^{5}+c_{1}y^{4}+d_{1}y^{3}+e_{1}y^{2}\nonumber \\
 & +f_{1}y+g_{1},\label{eq:best-fitx-1}\end{align}
 \begin{align}
x_{2} & =a_{2}y^{6}+b_{2}y^{5}+c_{2}y^{4}+d_{2}y^{3}+e_{2}y^{2}\nonumber \\
 & +f_{2}y+g_{2},\label{eq:best-fitx-2}\end{align}
 where $x_{1}$ is the polynomial curve fit for Fig. \ref{fig:n-phi-air},
$x_{2}$ is the polynomial curve fit for Fig. \ref{fig:n-theta-air},
and the coefficients are given by \begin{eqnarray*}
 & a_{1}=-8\times10^{13},\; b_{1}=2\times10^{11},\; c_{1}=-1\times10^{8},\\
 & d_{1}=52523,\; e_{1}=403.1,\; f_{1}=3\times10^{-4},\\
 & g_{1}=-3\times10^{-9},\end{eqnarray*}
 \begin{eqnarray*}
 & a_{2}=5\times10^{13},\; b_{2}=-1\times10^{11},\; c_{2}=1\times10^{8},\\
 & d_{2}=-29916,\; e_{2}=-264,\; f_{2}=-2\times10^{-4},\\
 & g_{2}=2\times10^{-9}.\end{eqnarray*}
  If $x_{1}$ represents a perfectly fitting polynomial functions
for the curve plotted in Fig. \ref{fig:n-phi-air}, then one should
expect the difference $x-x_{1}$ is a constant, where $x$ is the
plotted curve in Fig. \ref{fig:n-phi-air}. Similarly, if $x_{2}$
represents a perfectly fitting polynomial functions for the curve
plotted in Fig. \ref{fig:n-theta-air}, then one expects the difference
$x-x_{2}$ is a constant, assuming $x$ now is the plotted curve in
Fig. \ref{fig:n-theta-air}. Contrarily, if $x-x_{1}$ (or $x-x_{2}$)
is not a constant, then the polynomial $x_{1}$ (or $x_{2}$) cannot
be a perfectly fitting polynomial function for the curve plotted in
Fig. \ref{fig:n-phi-air} (or Fig. \ref{fig:n-theta-air}). %
\begin{figure}[H]
\begin{centering}
\includegraphics[scale=0.65]{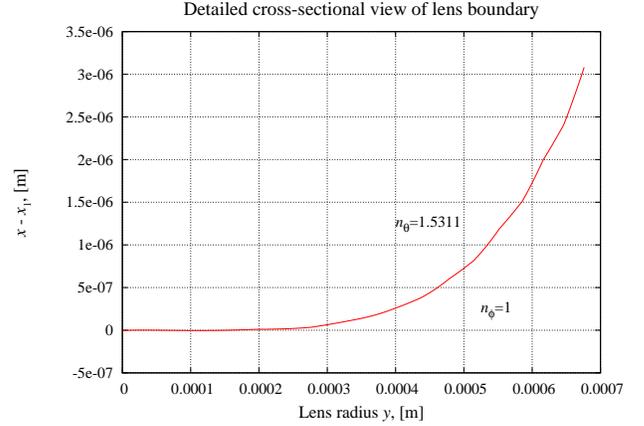}
\par\end{centering}

\caption{\label{fig:diff-phi-air} Plot of $x-x_{1}$ for the case where $n_{\phi}=1$
and $n_{\theta}=1.5311.$ }

\end{figure}
\begin{figure}[H]
\begin{centering}
\includegraphics[scale=0.65]{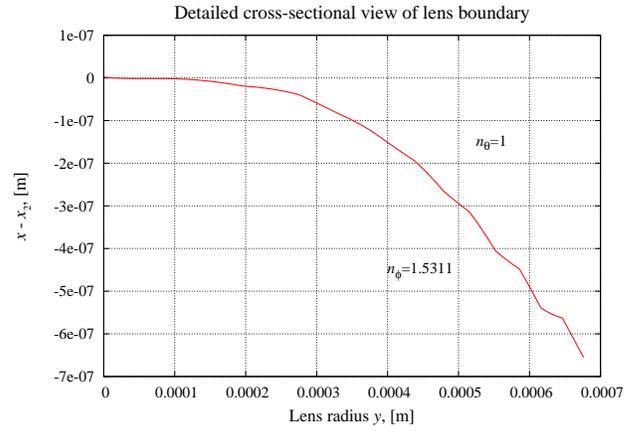}
\par\end{centering}

\caption{\label{fig:diff-theta-air} Plot of $x-x_{2}$ for the case where
$n_{\phi}=1.5311$ and $n_{\theta}=1.$}

\end{figure}
 The results are shown in Figs. \ref{fig:diff-phi-air} and \ref{fig:diff-theta-air}
respectively for the cases where $n_{\theta}>n_{\phi}$ and $n_{\theta}<n_{\phi}.$
It clearly shows that $x-x_{1}$ or $x-x_{2}$ are far from being
constants. This indicates that the surface cross-sectional profile
of lens is not a simple parabolic curve as deceptively portrayed Fig.
\ref{fig:n-phi-air} (or Fig. \ref{fig:n-theta-air}). Instead, the
magnification of the surface reveals series of kinked segments which
must be attributed to the discrete continuous curve segments in instantaneous
focal length $\left(\beta\right)$ versus the lens radius $\left(y\right)$
data shown in Fig. \ref{fig:lens-data-0}.

\section{Concluding Remarks}

At the heart of software assisted image focusing technology is the
specialized image reconstruction algorithm, which is permanently coded
into the built in digital signal processor. The algorithm is often
codified basing on the instantaneous focal length versus the lens
radius data as the initial input. The software assisted image focusing
system therefore requires a specially designed lens whose output generates
the instantaneous focal length versus the lens radius data. In this
work, a formula for the lens surface profile has been presented. The
derived lens formula generates a unique surface profile for the lens
based on the instantaneous focal length versus the lens radius data.
The lens design based on instantaneous focal length versus the lens
radius data makes this result well suited for software assisted image
focusing technology.

\section{Acknowledgments}

I would like to thank G. Alexander and A. Lukyanov for  providing
the raw data for their instantaneous focal length versus the lens
radius profile described in their patent. I would also like to thank
Dr. Seungwan Lee for the verification of the result using CODE V\textregistered{}%
\footnote{CODE V\textregistered{} is an optical design program with graphical
user interface for image forming and fiber optical systems by Optical
Research Associates (ORA), an organization that has been supporting
customer success for over 40 years (www.opticalres.com).%
} optical simulation tool. The author acknowledges the support for
this work provided by Samsung Electronics, Ltd.

\end{document}